\documentclass[preprint,9pt,sort&compress]{elsarticle}

\usepackage{amssymb}
\usepackage{epsfig}
\usepackage{graphics,graphicx}
\usepackage{subfigure}
\usepackage[nolist]{acronym}
\usepackage{lineno}
\usepackage{enumitem}
\usepackage{amsmath}

\makeatletter
\newcounter{RomanNumber}
\newcommand{\rom}[1]{\text{\setcounter{RomanNumber}{#1}\Roman{RomanNumber}}}
\newcommand{\coo}{\text{CO}_2}
\newcommand{\nn}{\text{N}_2}

\newcommand{\oC}{^{\circ}\text{C}}

\newcommand{\um}{\mu \text{m}}
\newcommand{\beq}{\begin{equation}}
\newcommand{\eeq}{\end{equation}}
\newcommand{\bitm}{\begin{itemize}}
\newcommand{\eitm}{\end{itemize}}
\makeatother

\journal{Nucl.Instrum.Meth.A}

\begin{document}

\begin{acronym}[Bash]

\acro{DEPFET} {DEPleted Field Effect Transistor}
\acro{SCB} {combined Support and Cooling Block}
\acro{FBG} {Fiber Bragg Grating}
\acro{2PACL} {2-Phase Accumulator controlled Loop}
\acro{FEA} {Finite Element Analysis}

\end{acronym}

\begin{frontmatter}

\title{Thermal mockup studies of the DEPFET pixel vertex detector for Belle $\rom{2}$}

\author{H.~Ye$^a$, C.~Niebuhr$^a$, R.~Stever$^a$, K.~Gadow$^a$, C.~Camien$^a$\\
(On behalf of the DEPFET Collaboration)
}

\address{
$^a$ \it\small DESY, Notkestrasse 85, D-22603 Hamburg, Germany
}

\begin{abstract}

The Belle $\rom{2}$ experiment currently under construction at the $e^+e^-$-collider SuperKEKB in Japan is designed to explore new physics beyond the standard model with an approximately 50 times larger data sample compared to its predecessor. The vertex detector (VXD), comprising a two layer DEPFET pixel detector (PXD) surrounded by four layers of double sided silicon strip detector (SVD), is indispensable for the accurate determination of the decay point of $B$ or $D$ meson as well as track reconstruction of low momentum particles. The DEPFET sensors in Belle $\rom{2}$ are thinned down to 75~$\um$ with low power consumption and low intrinsic noise. In the DEPFET concept, the front-end electronics is placed outside of the sensitive area, and thus no cooling components are necessary inside the physics acceptance of the detector.
Evaporative two-phase CO$_2$ cooling in combination with forced air flow has been chosen as the scheme for the PXD cooling.
To guarantee the DEPFET detector operation condition and verify the cooling concept, a PXD mockup is constructed at DESY. Studies of the thermal and mechanical performance are presented in this paper.

\end{abstract}


\end{frontmatter}



\section{Introduction}
\label{intro}

As an upgrade of the asymmetric electron-positron collider KEKB, SuperKEKB~\cite{superkekb} at KEK in Tsukuba, Japan aims at increasing the peak luminosity by a factor of 40 to 8$\times$10$^{35}$~cm$^{-2}$s$^{-1}$.
Belle $\rom{2}$~\cite{tdr} is an extensive upgrade of the former Belle experiment with the goal of performing high-precision measurements of rare decays, to explore new physics beyond the Standard Model at the intensity frontier. Commissioning of the accelerator with the detector still in parking position started in February 2016. The full experiment is scheduled to begin operation in 2018. Belle $\rom{2}$ is expected to accumulate an integrated luminosity of about 50~ab$^{-1}$ well within the next decade.

\begin{figure}[htbp]
\centering
\includegraphics[angle=0,width=0.5\textwidth]{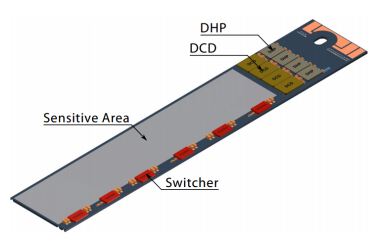}
\caption{Illustration of the DEPFET sensor with readout electronics and thinned sensitive area for the outer layer of the Belle $\rom{2}$ PXD. The front-end electronics of Switchers are placed at the rim of the sensor, while the DCD and DHP chips are located outside the physics acceptance.}
\label{module}
\end{figure}

The innermost part of the Belle $\rom{2}$ vertex detector (VXD) is a two-layer highly granulated pixel detector (PXD)~\cite{pxd}, surrounded by a four-layer double sided silicon strip vertex detector (SVD)~\cite{svd}.
The Belle $\rom{2}$ PXD employs the novel semiconductor detector concept of \ac{DEPFET}~\cite{depfet}.
This technology is an attractive choice for  vertex detectors in particle physics experiments due to its low material budget, excellent noise performance, low power consumption and high spatial resolution.
The DEPFET concept combines particle detection and signal amplification in one device by embedding the FETs into a fully depleted silicon bulk. It allows very thin sensors. The thickness of the Belle $\rom{2}$ PXD sensor (Fig.~\ref{module}) has been chosen as  75~$\um$, the matrix comprises 250$\times$768 pixels with dimensions of 50$\times$60(75)~$\um$.
The PXD is read out in so called rolling shutter mode. Each row of pixels is selected by pulling the gate line to a negative potential using the steering chip (Switcher) that is placed at the rim of the sensor. The selected pixels send currents down the vertically connected drain lines. These currents are processed at the bottom of the matrix by the Drain Current Digitizer (DCD) chips, which perform an immediate digitization and send data to the Data Handling Processor (DHP), which buffers and analyzes the digital data stream and performs a zero suppression.
A key feature of the DEPFET sensor design is, that most of the heat load occurs outside the physics acceptance at the end-of-stave (EOS), where the 4 DCD and DHP chips together dissipate 8~W per sensor. The thin sensors and the Switcher chips generate only very little extra heat ($\sim 1$W) which can be removed by forced air flow in the acceptance region.

For optimal performance a sufficiently low and homogeneous temperature distribution is required on the DEPFET sensors, detailed cooling requirements are described in Sec.~\ref{coolingconcept}.
In order to verify and optimize the cooling concept, a full sized PXD mockup with the same mechanical and thermal properties as the final detector was built at DESY.
Measurements with this mockup that yield valuable insights into the operation of the DEPFET vertex detector are presented in this paper.

\begin{figure}[htbp]
\centering
\includegraphics[angle=0,width=0.75\textwidth]{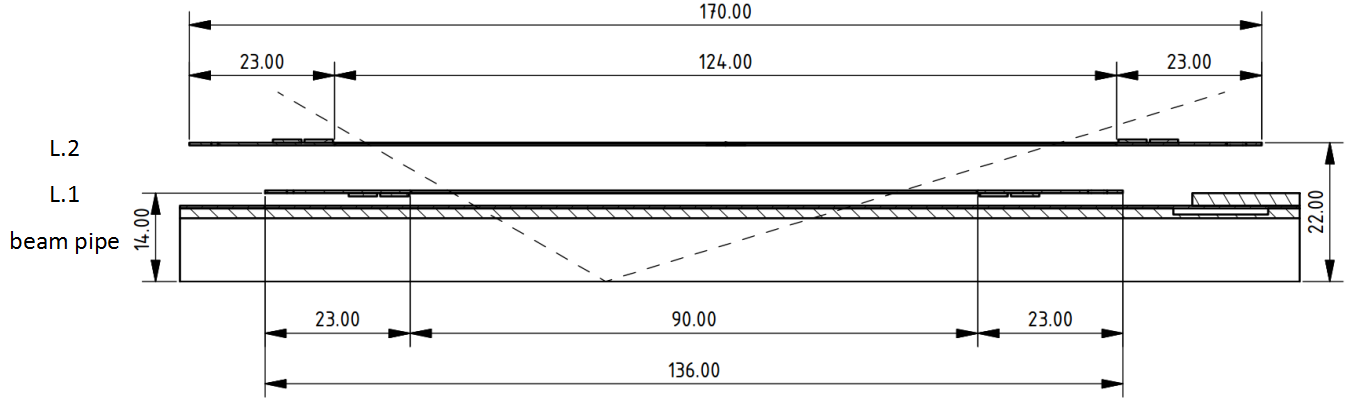}
\caption{Cross sectional view along the beam line of the Belle $\rom{2}$ PXD detector and the central beam pipe. The PXD consists of two layers DEPFET pixels at the radii of 14 and 22~mm. A pair of mirrored DEPFET sensors, which are glued together end-to-end, are named ladder. The inner (outer) layer, denoted as L.1 (L.2), is composed of 8 (12) PXD ladders.
}
\label{layout1}
\end{figure}

\begin{figure}[htbp]
\centering
\includegraphics[angle=0,width=0.75\textwidth]{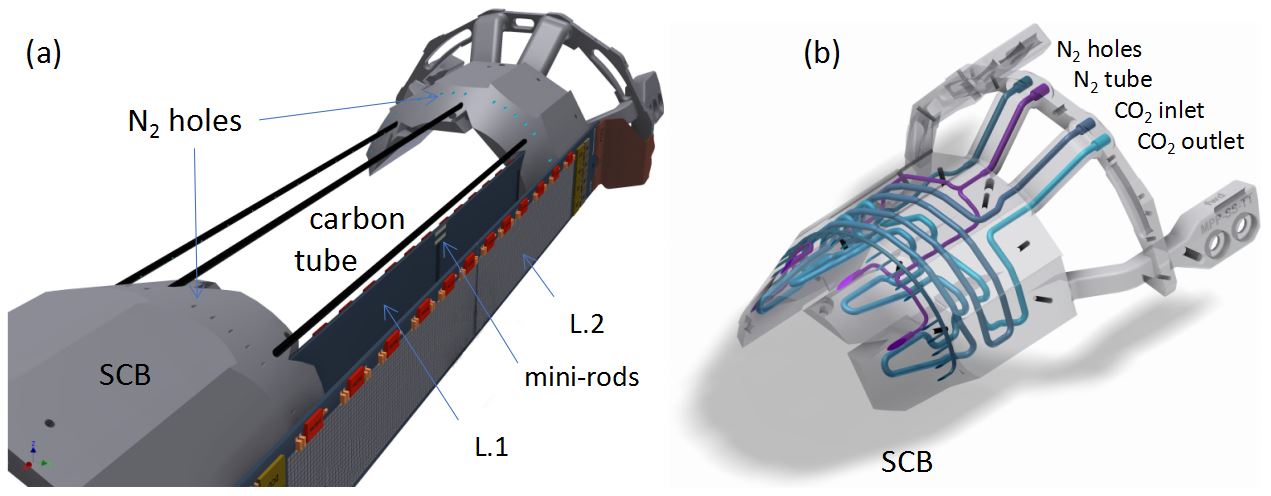}
\caption{(a) The mechanical design of the Belle $\rom{2}$ PXD. The PXD ladders are fixed on two pairs of \ac{SCB}s, which are mounted on the central beam pipe. Forward and backward SCBs are connected with 8 carbon fiber tubes for $\nn$ cooling of the Switcher chips of the inner layer.
(b) 3D printed SCB with integrated cooling channels for $\coo$ circulation and open channels providing forced $\nn$ flow.
}
\label{scb}
\end{figure}

\begin{figure}[htbp]
\centering
\includegraphics[angle=0,width=0.75\textwidth]{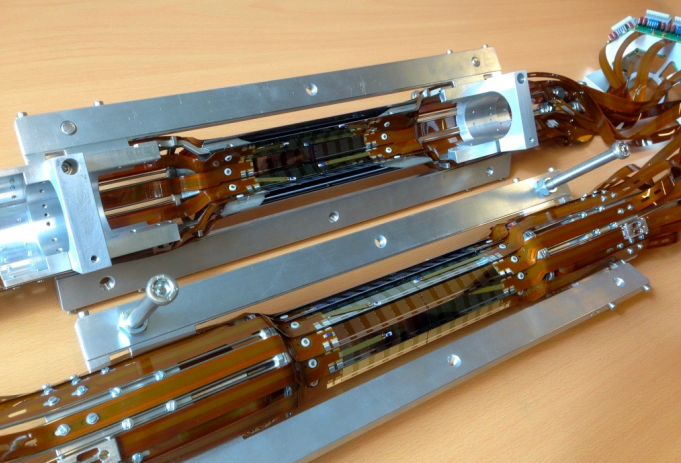}
\caption{Assembly of PXD thermal dummy sensors at DESY. The sensors are screwed on the SCBs. After mounting the two half shells on the dummy beam pipe the support bars on the sides are removed.}
\label{assembly}
\end{figure}

\section{Mechanical Layout and Cooling Concept of PXD}
\label{coolingconcept}

The two layers DEPFET PXD sensors are placed very closely to the beam pipe for a optimal vertex reconstruction. The inner (outer) layer, denoted as L.1(L.2), is composed of 8(12) ladders. Each ladder, is formed from a pair of mirrored DEPFET sensors, which are glued together end-to-end and combined with additional ceramic mini-rods embedded in the thick rim of sensor. Such a mechanical structure makes the DEPFET sensors self-supporting with no necessity of further holding structures in the sensitive region. The sensor locating at lower z coordinate is named backward type while the other named forward type. The total dimensions for L.1(L.2) are 136.00(170.00) $\times$15.40$\times$0.525~mm$^3$ with the sensitive area of 89.60(122.88)$\times$12.50~mm$^2$, as sketched in Fig.~\ref{layout1}.

Two rings of \ac{SCB}s are used to mount PXD onto the beam pipe. Each ring is split into two identical components (Fig.~\ref{scb}). The steely SCBs are manufactured using 3D printing technology, which allows the integration of complicated closed CO$_2$ cooling channels to cool the end of the sensors, and open nitrogen channels with a diameter of 0.3mm to provide airflow~\cite{pablo}.
The slanted part of the forward SCB accommodates a lower angle due to the asymmetric Belle $\rom{2}$ acceptance, and needs a slightly different layout of the integrated channels. In total 8 carbon tubes with an outer diameter of 1mm connect the backward and forward SCBs. With the precision holes along the tubes, airflow is provided to cool the Switchers on L.1. The tubes are coated with silver paint for grounding purposes.
Both layers of PXD ladders are mounted on the common \ac{SCB} and fixed by M1.2 screws with plastic washers.
Elongated holes for the screws at the end of the forward sensor are adopted to allow possible shifts in longitudinal direction due to thermal expansion.

Adequate cooling is required in such a dense layout to achieve a controlled temperature  environment for PXD to minimize the shot noise resulting from leakage current.
For Belle $\rom{2}$ PXD, the temperature on the sensor is required to be below 25$\oC$, and on the chips to stay below $\sim 50 \oC$ to avoid the risk of electro-migration. The total power in the PXD system is about 360~W, of which 320~W are dissipated by DCD/DHP, and 20~W each by sensors and Switchers.

The large heat load at DCD and DHP will increase the local temperature by 40$\oC$, causing a higher pressure drop in the CO$_2$ cooling channels.
The CO$_2$ mass flow of 1~g/s is required in each cooling circuit, to do effective cooling and defend the risk of dry-out\footnote{The dry-out phenomenon happens when the mass flow is so low that no liquid film on the pipe walls, the heat transfer is strongly reduced and results in sudden temperature increase~\cite{co2_2}.} but not results large pressure drop in the circuit.
The \ac{FEA} establishes such amount of dissipated heat requires the $\coo$ temperature to be lower than -20$\oC$ .

The cooling system for the Belle $\rom{2}$ VXD is based on the \ac{2PACL} method~\cite{co2,co2_2}, which has been originally developed for the thermal control system of the AMS tracker~\cite{amstracker}, and was later implemented also in the LHCb experiment for the cooling of the VErtex LOcator (VELO)~\cite{velo}.
Meanwhile evaporative two phase CO$_2$ cooling has been generally adopted as cooling concept for vertex detectors due to the extreme radiation hardness of the cooling agent and the possibility to use small diameter cooling tubes resulting in relatively small amount of extra dead material.
As refrigerant, CO$_2$ provides a high heat transfer coefficients that is one order of magnitude higher than traditional refrigerants, together with relatively high evaporation pressure ($\sim$ 55~bar at 20$\oC$).
In the two-phase regime the heat removal is achieved by evaporating liquid CO$_2$ at constant temperature and pressure.
The 2PACL pumped system uses a 2-phase accumulator, that can be both heated and cooled, to control the pressure and hence temperature inside the detector volume. The cooling plant can be placed far away from the detector where it can easily be maintained and only small diameter cooling tubing is necessary inside the detector.

\begin{figure}[htbp]
\centering
\includegraphics[angle=0,width=0.6\textwidth]{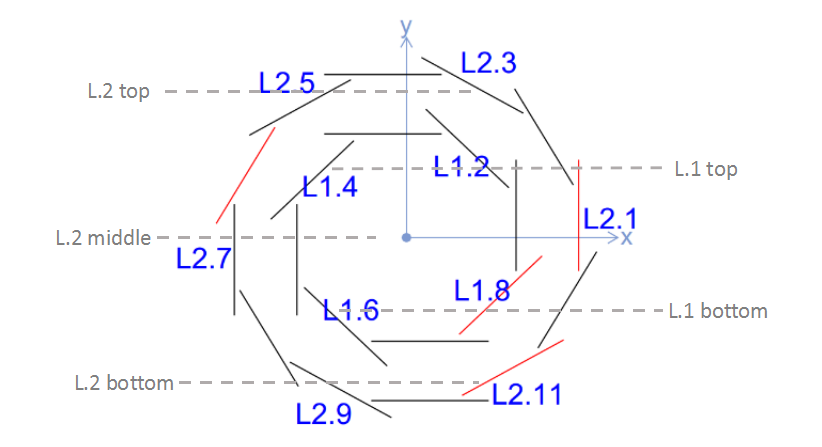}
\caption{Naming scheme of Belle $\rom{2}$ PXD. In the shortcut of L.{\it a.b}, $a=1,2$ denotes the layer number, and $b$ denotes the ladder number, which goes from 1 to 8 (12) for L.1 (L.2). The ladders with Pt100s glued are labeled. Thermal performance of the ladders marked in red is monitored with the infrared camera. The dashed lines indicate the y-slices of L.1(2)-top(/middle)/bottom in Figs.~\ref{tdist}, \ref{tco2}, \ref{tn2flow} legends. }
\label{layout}
\end{figure}

\section{Experimental setup}

For the thermal mockup, dummy sensors (Fig.~\ref{assembly}) have been manufactured exactly in the same way as for the final detector ~\cite{andricek}.
Instead of real electronics, the resistive dummy loads are used to simulate the power distribution in the functioning ladder. The "sensor", "Switcher" and "DCD/DHP"-like resistors with the nominal power load of 0.5~W, 0.5~W and 8~W are integrated in each half ladder\footnote{the power dissipation are based on the initial numbers for the first versions of chips for DEPFET, the final numbers need to be confirmed in future.}.
An extra power of 25~W is given on the kapton cables to simulate their power dissipation.
The heat intake arising from the SuperKEKB beam pipe and SVD are not taken into consideration.

\begin{figure}[htbp]
\centering
\includegraphics[angle=0,width=0.75\textwidth]{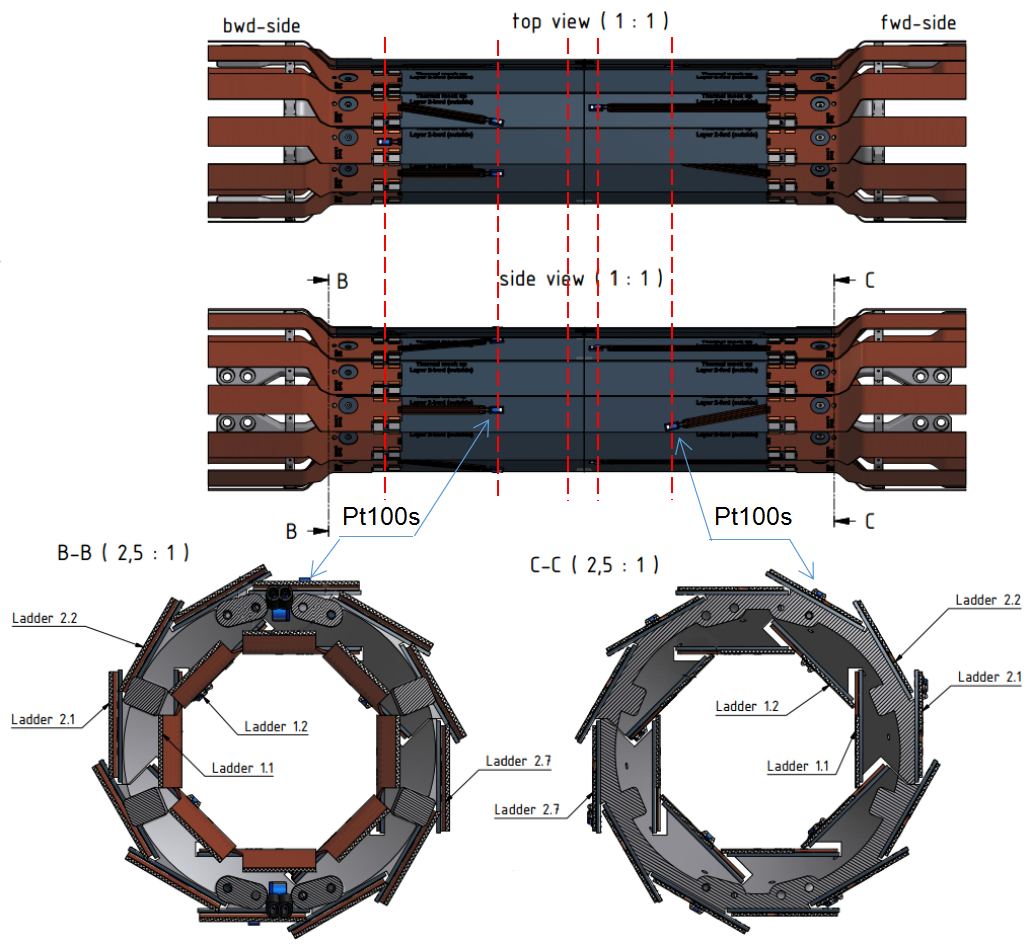}
\caption{The position of Pt100s in the PXD mockup.}
\label{pt100s}
\end{figure}

The M1.2 screws of both PXD layers are fixed with a torque of about 0.07Nm onto the SCBs.
Too little torque will lead to insufficient thermal contact resulting in too hot ASICs, while too large torque may block the shift at the long hole and lead to detectable deformation of the ladders, furthermore increases the risk to break the silicon underneath the washers. The SCB surfaces contacted the sensors were not specially treated in the thermal mockup.

20 Pt100 resistance thermometers are glued at 5 different z-slices along the ladders to probe the local temperature. In each z-slice there are at most 4 Pt100s locating at different (x,y) coordinates, either on the sensitive area or near the Switchers, as shown in Fig.~\ref{pt100s}.
The 4 polyimide coated fiber optical sensors (FOS), each with 4 \ac{FBG} inscribed~\cite{fbg}, are positioned a few millimeters offset from L.2.5 or L.2.11 (ladders are denoted as L$.a.b$, where $a$ is layer number and $b$ is ladder number, as illustrated in Fig.~\ref{layout}) to monitor the temperature and humidity in the volume. Three of them are protected with a Teflon hermetic tube to suppress the sensitivity to humidity, and thus offer compensation to the forth FOS for humidity monitoring.
A thermal imaging infrared camera is also used to measure the temperature distribution along the ladders.

The whole PXD thermal mockup was kept in a plastic cylinder (inner diameter of 18~cm, length of 70~cm) serving as dry volume. N$_2$ is injected into the volume through the SCB open channels, to generate a forced convection and avoid condensation. When the fully heat loaded PXD operated with $\coo$ set point of -30~$\oC$ and N$_2$ flow of 23~L/min, the dew point inside the volume was reduced to about -35~$\oC$, and the humidity decreased to be less than 5\%, measured by the FOS.

\section{Thermal Studies}

\begin{figure}[htbp]
\centering
\includegraphics[angle=0,width=0.8\textwidth]{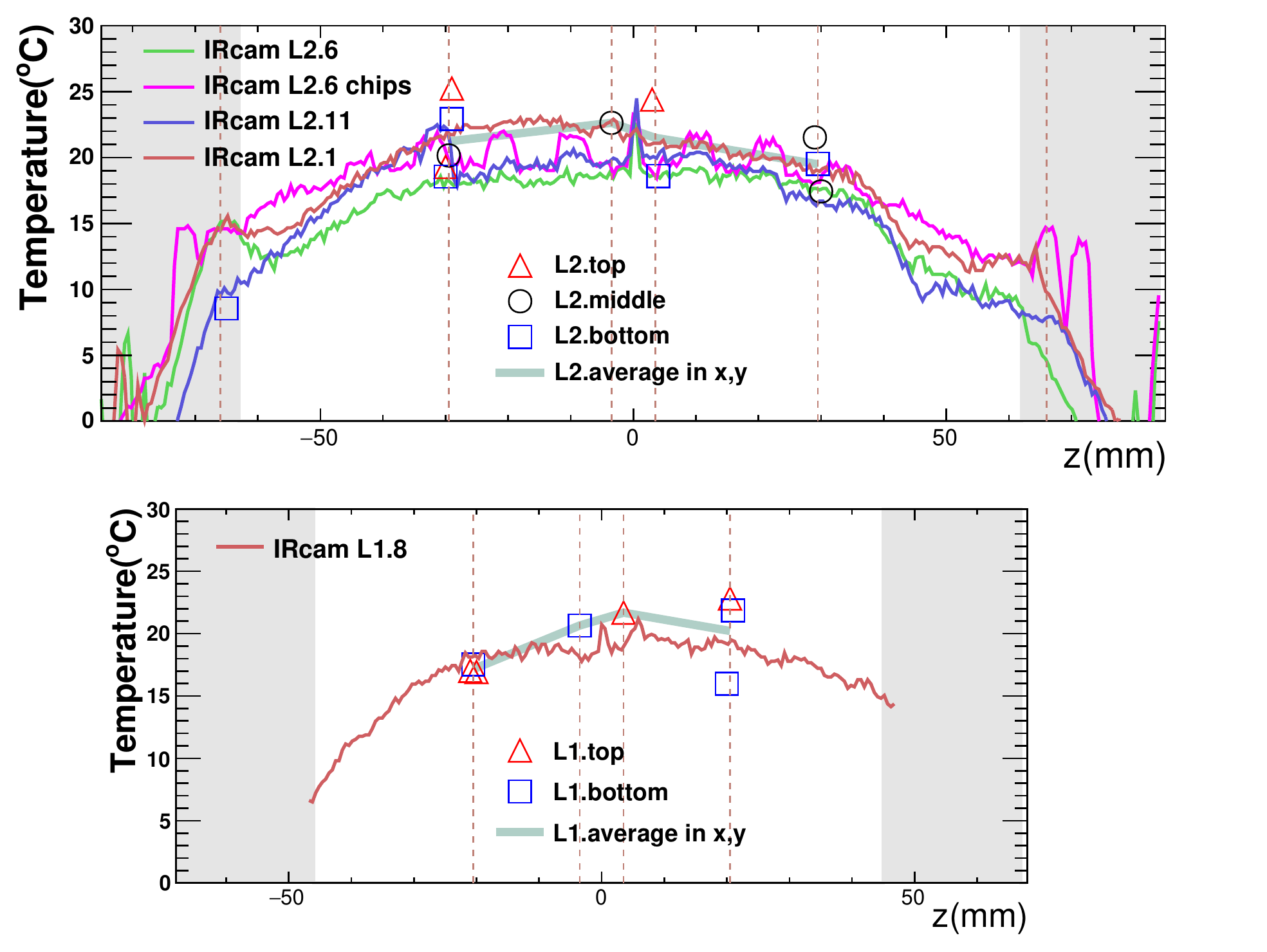}
\caption{The longitudinal temperature distribution of PXD ladders along the z-direction. The top (bottom) plot shows the L.2(1). In each plot the gray areas indicate the regions of DCD/DHP, while the 75~$\mu$m thick sensitive area is shown in the center. The markers show the temperature measured by Pt100s at certain position along z-axis. Different marker sharps indicate the different y positions of Pt100s, as illustrated in Fig.~\ref{layout}. The same markers in the identical z-slice are slightly shifted for better visibility. The thick solid lines indicate the temperature averaged in (x,y), measured from the Pt100s. The thin solid lines show the temperature distribution measured by the IR camera on selected ladders, which is marked in red on Fig.~\ref{layout}.
}
\label{tdist}
\end{figure}

Figure~\ref{tdist} shows the longitudinal temperature distribution measured on a fully heat-loaded PXD ladder with the $\coo$ temperature set to -30$\oC$ with $\nn$ injected with a flow of 23~L/min at room temperature. The ladders were efficiently cooled with such a  combined cooling scheme, achieving a temperature of 10$\oC$ at the highly loaded DCD/DHP chips, and between 15 and 25$\oC$ in the sensitive region. The temperature gradient along the sensitive area was kept at about 5$\oC$ for the entire length with convective and active cooling. Due to the higher density of cold N$_2$, a top-bottom temperature gradient of about 5$\oC$ for L.2 was measured by the Pt100s. The Switchers temperature was about 2-3$\oC$ higher than at the sensitive area, indicated by infrared camera.
Temperature measured by the 3 protected FOS were congruent to the results of Pt100s.

\subsection{CO$_2$ Cooling}

\begin{figure}[htbp]
\centering
\includegraphics[angle=0,width=0.75\textwidth]{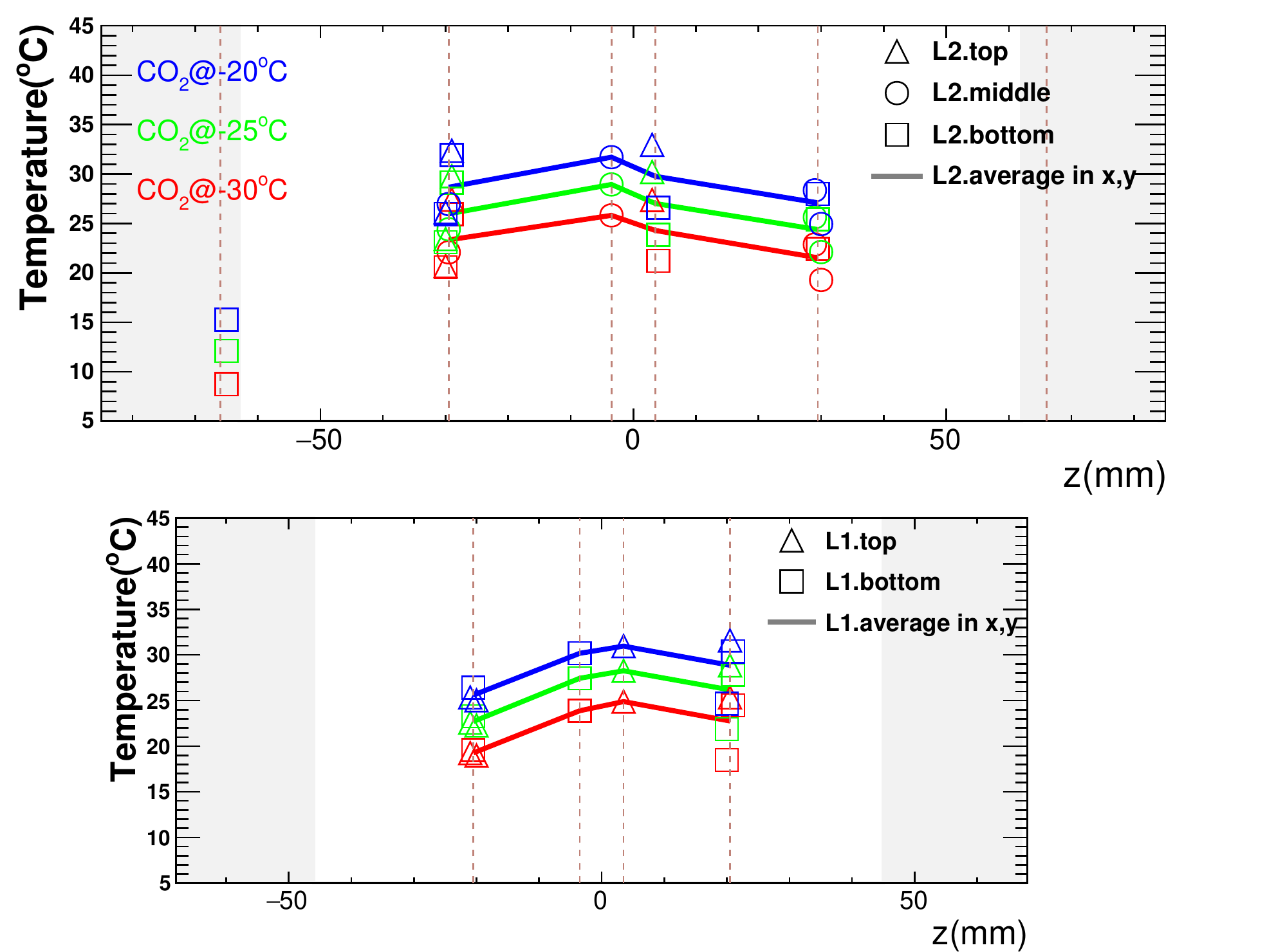}
\caption{The temperature distribution on PXD sensors at different CO$_2$ temperature set points. The N$_2$ flow of 20L/min is injected.}
\label{tco2}
\end{figure}

The temperature distribution in PXD at different CO$_2$ set temperature are compared in Fig.~\ref{tco2}. With the identical N$_2$ injection of 20~L/min, every 5$\oC$'s change in CO$_2$ temperature shifted the temperature distribution of the system by about 3$\oC$, while the gradient along the ladder stayed.

\subsection{N$_2$ Cooling}

\begin{figure}[htbp]
\centering
\includegraphics[angle=0,width=0.75\textwidth]{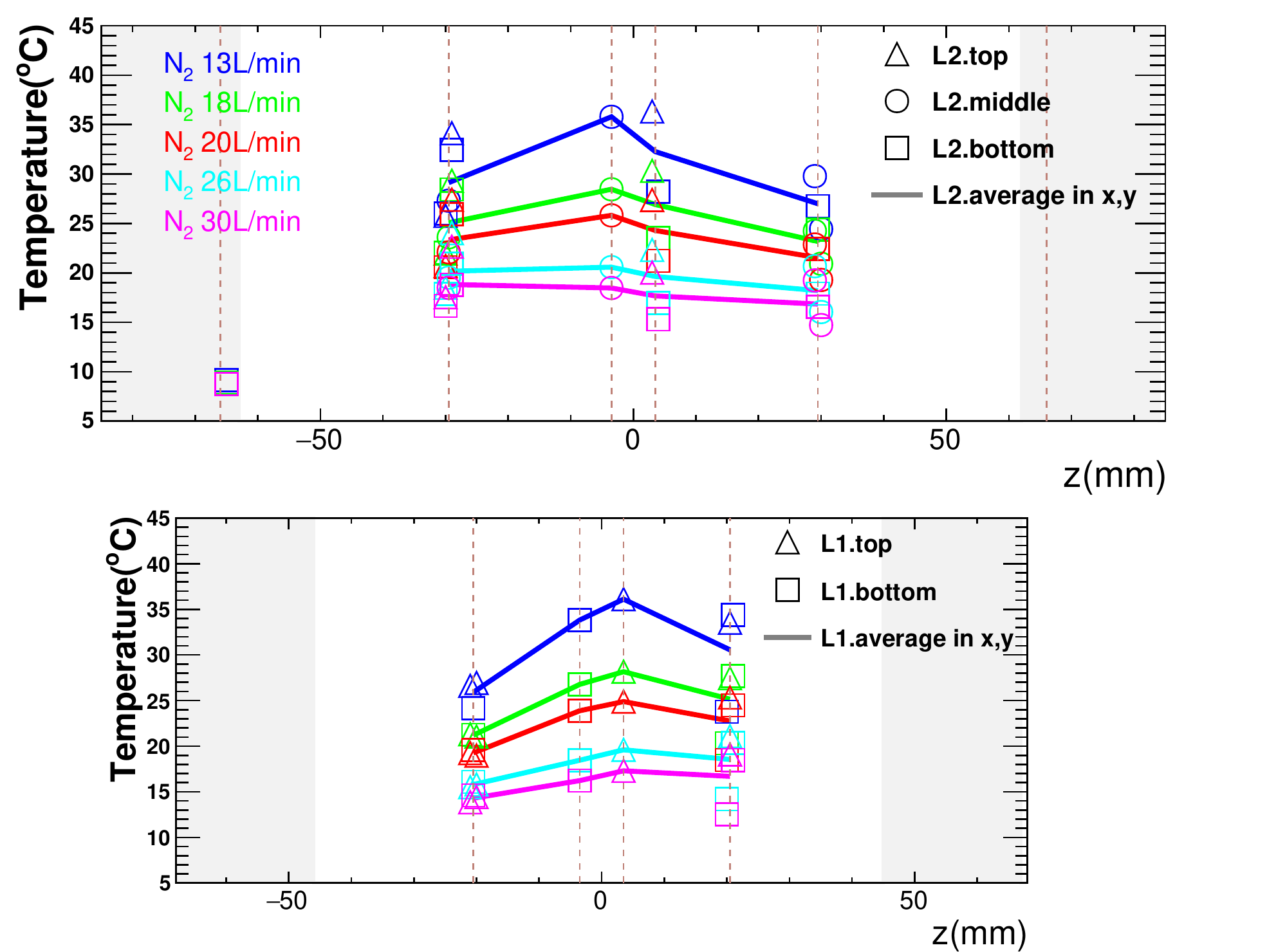}
\caption{The temperature distribution on PXD sensors at different N$_2$ flow. The CO$_2$ temperature is set at -30$\oC$.}
\label{tn2flow}
\end{figure}

The temperature distribution in PXD as a function of the injected N$_2$ flow is presented in Fig.~\ref{tn2flow}. Within the identical set point of T(CO$_2$), the higher N$_2$ flow would result in a lower gradient along the sensor. The N$_2$ injected from the carbon tubes is critical for L.1, and the N$_2$ open channels on SCBs are critical for L.2. The combination of both N$_2$ injection gave good cooling performance along the sensors.

The effect on the performance of varying the temperature of the injected $\nn$ was also studied.
\ac{FEA} simulations indicate that after about 0.5~m of unprotected pipe length, the N$_2$ will assume  ambient temperature.
In the real detector the unprotected N$_2$ lines will pass  through a region of about 35$\oC$.
To simulate the influence of these hot N$_2$ lines, we introduce 44W of external heat to N$_2$ lines before they reach the SCBs, resulting in a local temperature on the N$_2$ lines of more than 40$\oC$. The measurements show that these hot N$_2$ lines only cause a moderate temperature increase of about 1$\oC$ in the system, indicating a very efficient cooling of the injected N$_2$ by the SCBs.


\subsection{Vibration}

\begin{figure}[htbp]
\centering
\includegraphics[angle=0,width=0.65\textwidth]{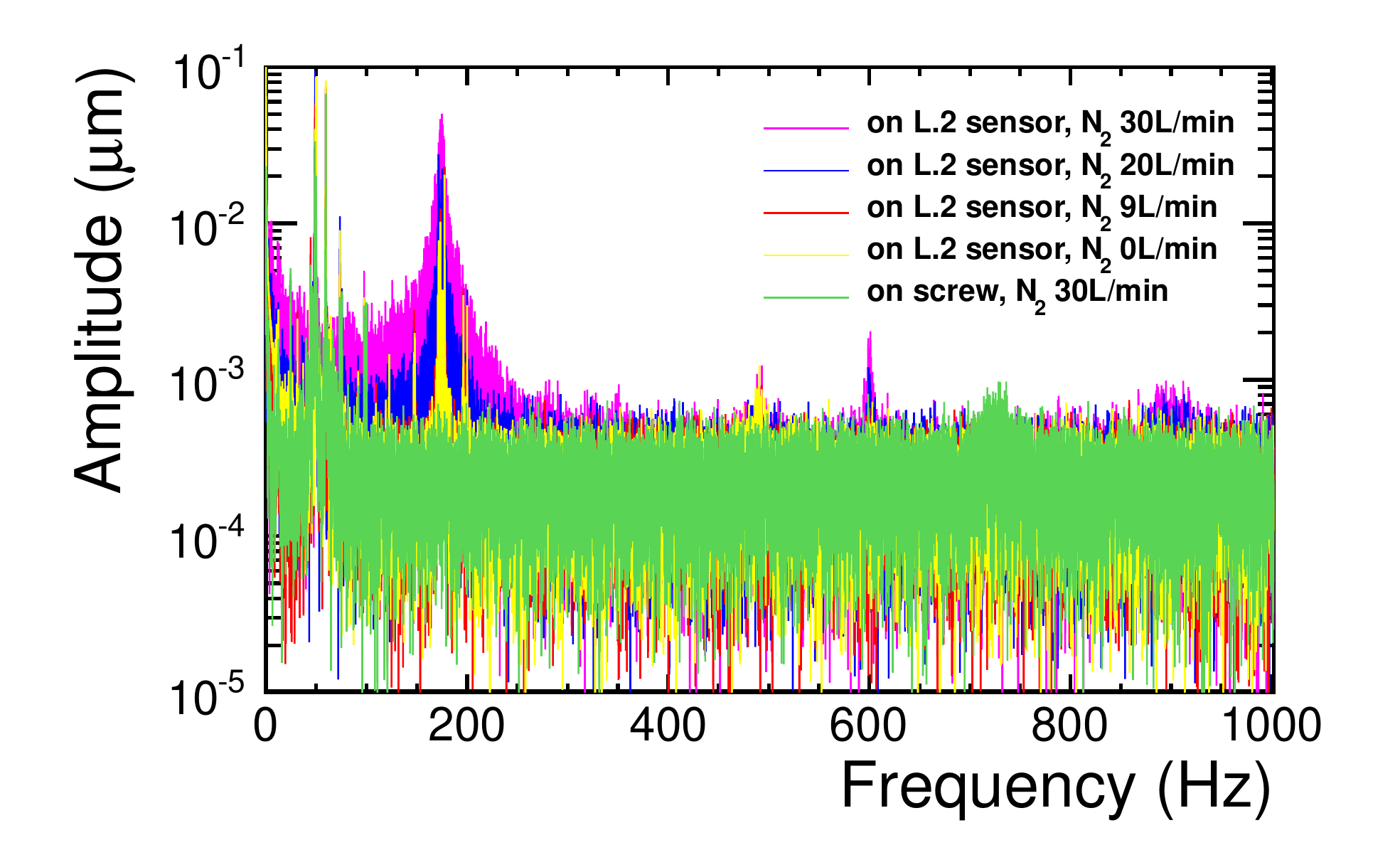}
\caption{Vibration in the center of L.2 ladder introduced by the gas injection, at different injected flow rate. The measurement on the SCB screw shows the background.}
\label{vibf5}
\end{figure}

Possible vibration of the sensors of the outer layer caused by the $\nn$ injection was monitored with the non-contact laser displacement sensor.
Fig.~\ref{vibf5} shows the Fourier spectrum of the vibration amplitude at the center of the ladder for different N$_2$ flows. A resonance at about 175~Hz was observed on a flat background with an amplitude of $\approx 2 \cdot 10^{-4}~\um$, as measured at the fixation screws on the SCB.
The vibration amplitude increased with the flow rate, it reached a level of about 0.02~$\um$ when 20~L/min of N$_2$ was injected.
The strong peaking background at 50~Hz raised from the nearby electronics in the lab.
Besides the vibration, a displacement of 0.5 $\mu$m due to the internal pressure was observed in the center of ladders of outer layer, at the N$_2$ flow of 20~L/min. Comparing to the spatial resolution of VXD and mechanical stability tolerance of the DEPFET sensor, the vibration amplitude and displacement is negligible in the experiment.

\section{Conclusion and Discussion}

The Belle $\rom{2}$ PXD employs the ultra low mass DEPFET pixels, the sensor is thinned down to 75~$\um$ and the high consumptive readout electronics can be placed outside the physics acceptance of detector. The self-supporting DEPFET ladders are mounted on the SCBs with integrated channels for the combined cooling concept of evaporative two phase CO$_2$ and forced $\nn$ flow.
To verify the mechanical and thermal performance of Belle $\rom{2}$ PXD, a full sized mockup is built with the same properties as the DEPFET pixel detector.The measurements based on the power consumption of the first version of chips are presented.
With the set point of CO$_2$ at -30$\oC$ and N$_2$ flow of 23~L/min, the temperature along the full loading PXD ladder stays less than 25$^{\circ}$C with a gradient of about 5$^{\circ}$C, according to the power dissipation of the initial version of readout chips. In the vertical direction a gradient of about 5$^{\circ}$C across the PXD volume results from the higher density of cold N$_2$. Cooling performance at different CO$_2$ set points and N$_2$ flow are studied. The temperature decreases with lower CO$_2$ temperature set point, the gradient decreases with larger N$_2$ flow. At the selected air flow rate a vibration with a frequency of 175~Hz and an amplitude of 0.02~$\mu$m as well as a deformation of 0.5~$\mu$m are introduced by the air cooling.

The DEPFET technology is a promising concept for the pixel detector in future particle physics experiments. Various thermal and mechanical measurements on this mockup can provide insights into the operation and cooling concept of DEPFET pixels. For Belle $\rom{2}$ PXD, the $\coo$ temperature is suggested to set at $\sim$ -30~$\oC$ and the $\nn$ flow of $\gtrsim $ 20~L/min. Although the goal is to make the mockup as realistic as possible, it can not fully reproduce the thermal performance of PXD. The intake heat load from the beam pipe and surrounding SVD sensors and cooling components are not included, and the Belle $\rom{2}$ VXD cylinder has larger and asymmetric volume. The thermal mockup studies verify the combined cooling method is effective for the PXD operation. The $\coo$ set point and $\nn$ flow can be further optimized when the Belle $\rom{2}$ experiment starts.

\section*{Acknowledgements}

We wish to thank the MPI f\"ur Physik, M\"unchen group for designing and preparing some cooling components including the SCB, we also thank the MPG Halbleiterlabor (HLL) group for producing the PXD dummy sensors and CSIC-UC group for preparing the FOS.
The Belle $\rom{2}$ VXD cooling framework has been developed on the basis of the experience gained with the ATLAS-IBL cooling system. We would like to acknowledge the support from CERN experts.


\bibliographystyle{elsarticle-num}
\bibliography{pxd_thermal_mockup}

\end{document}